\documentclass[prb,twocolumn,showpacs]{revtex4-1}
\usepackage{epsfig}
\usepackage{graphicx}
\usepackage{amsfonts,amssymb}
\usepackage{amsmath}
\usepackage{color}

\newcommand{\be}{\begin{equation}}
\newcommand{\ee}{\end{equation}}   
\newcommand{\bea}{\begin{eqnarray}}
\newcommand{\eea}{\end{eqnarray}}
\newcommand{\ba}{\begin{array}}
\newcommand{\ea}{\end{array}}

\newcommand{\q}{{\bf q}}
\renewcommand{\k}{{\bf k}}
\newcommand{\Q}{{\bf Q}}

\begin{document}
\title{Two-orbital model for CeB$_6$}
\author{Dheeraj Kumar Singh}
\email{dheerajsingh@hri.res.in} 
 \affiliation{Harish-Chandra Research Institute, Chhatnag Road, Jhunsi, Allahabad 211019, India}
\begin{abstract}
We describe a two-orbital tight-binding model with bases belonging to the $\Gamma_8$ 
quartet. It captures several characteristics of the Fermiology unravelled by the recent 
angle-resolved photoemission spectroscopic (ARPES) measurements on cerium hexaboride CeB$_6$ samples cleaved 
along different high-symmetry crystallographic directions, which includes the 
ellipsoid-like Fermi surfaces (FSs) with major
axes directed along $\Gamma$-X. We calculate various multipolar susceptibilities 
within the model and identify the susceptibility that shows the strongest 
divergence in the presence of standard onsite Coulomb interactions and discuss it's possible 
implication and relevance with regard to the signature of strong ferromagnetic correlations 
existent in various phases as shown by the recent experiments.
\end{abstract}

\maketitle
\newpage
\section{Introduction}
Strongly correlated $f$-electron systems exhibit a wide range of ordering phenomena including various
magnetic orderings as well as superconductivity.\cite{mydosh,thalmeir1} However, they are notorious for 
possessing complex ordered phases or so called 'hidden order', which are sometimes
not easily accessible experimentally because of the ordering of multipoles 
of higher rank such as electric quadrupolar, magnetic octupole etc. than the rank one magnetic 
dipole.\cite{kuramoto,santini} 
This marked difference from the correlated $d$-electron system is a 
result of otherwise a strong spin-orbit coupling existent in these systems. Recent predictions of 
samarium hexaboride (SmB$_6$) to be a topological Kondo insulator has led to an 
intense interest and activities in these materials.\cite{takimoto2}

CeB$_6$ with a simple cubic crystal structure is one of the most extensively studied $f$-electron 
system both theoretically as well as experimentally. Apart from the pronounced Kondo 
lattice properties, it undergoes two different types of ordering transition as a function of temperature despite 
it's simple crystal structure.\cite{kasuya1} First, there is a transition to the AFQ phase with ordering wavevector
$\Q_1 = (\pi, \pi, \pi)$ at $T_{\Q} \approx 3.2K$, which has long remained
hidden to the standard experimental probes such as neutron diffraction.\cite{effantin,goodrich,matsumura,shiina,thalmeier}
Then, another transition to the AFM phase with double $\Q_2$ commensurate structure with 
$\Q_2 = (\pi/2, \pi/2, 0)$ takes place at $T_{N} \approx 2.3K$.\cite{zaharko}

{Significant progress has been made recently through the experiments in understanding the
nature of above mentioned phases of CeB$_6$. Magnetic spin resonance, for instance, has been observed in 
the AFQ phase\cite{demishev1,demishev2} with it's 
origin attributed to the ferromagnetic correlations\cite{krellner,schlottmann} as 
in the Yb compounds, e.g., YbRh,\cite{krellner} YbIr$_2$Si$_2$,\cite{sichelschmidt}
and one Ce compound CeRuPO.\cite{bruning} On the other hand, according to a recent 
inelastic neutron-scattering (INS) experiment, AFM phase is rather a 
coexistence phase consisting of AFQ ordering as well.\cite{friemel} In another INS measurements,
low-enengy ferromagnetic fluctuations have been reported to be more
intense than the mode corresponding to the magnetic ordering wavevector $\Q_2$ in the AFM phase, 
which stays though with reduced intensity even in the pure AFQ phase.\cite{jang} Overall picture emerging 
from these experiments and hotspot observed near $\Gamma$ by ARPES imply the existence of 
strong ferromagnetic fluctuations in various phases of CeB$_6$.

So far most of the theoretical studies have focused on the localized aspects of 4-$f$ electron while 
neglecting the itinerant character when investigating multipole orderings.\cite{shiina,thalmeier,okhawa} However,
this may appear surprising because the estimates of density of 
states (DOS) for CeB$_6$ at the Fermi level from low-temperature specific heat measurement as well as from 
the effective mass measurement from de Haas-van Alphen (dHvA) gives a significantly larger value when compared to the 
paramagnetic metal such as LaB$_6$ provided that the FSs are considered same in both the compounds.\cite{harrison} In the 
temperature regime $T > T_{Q}$, it exhibits a typical dense Kondo behavior dominated by Fermi liquid 
with a Kondo temperature of 
the order of $T_N$ and $T_{Q}$.\cite{nakamura} Moreover, a low energy dispersionless collective mode 
at $\Q_1$ has been observed in the INS experiments, which is well within the single particle charge 
gap present in the coexistence phase.\cite{friemel} The existence of such spin excitons have been reported in 
several superconductors\cite{eremin} as well as heavy-ferimion compounds\cite{akbari} previously, and 
explanation for the origin of such modes has been provided in terms of 
correlated partilce-hole excitation a characteristics of the itinerant systems. 

Recent advancement based on a full 3D tomographic sampling of 
the electronic structure by the APRES has unraveled  
the FSs in the high-symmetry planes of cubic CeB$_6$.\cite{koitzsch,neupane} FSs are 
found to be the cross sections of the 
ellipsoids, which exclude the $\Gamma$ point and are bisected by (100) plane at $k_z = \pi$. The largest
semi-principle axes of the ellipsoid coincides with $\Gamma$-X. Based on the FS characteristics, it 
has been suggested that multipole order may arise due to the nesting as the shifting of one ellipsoid by 
nesting vector ($\pi, \pi, \pi$) into the void formed in between other three can
result in a significant overlap. Interestingly, the features of FS bear 
several similarities to those of LaB$_6$, which has also been suggested by earlier 
estimates based mainly on the dHvA experiments\cite{onuki,harrison} as well as by several band-structure 
calculations.\cite{kasuya,suvasini,auluck} 

Despite various experimental works on the FSs of CeB$_6$, no theoretical studies of 
ordering phenomena have been carried out within the models based on the 
realistic electronic structure, and therefore the nature of instability or 
fluctuations that will arise in that case is of strong current
interest. To address this important issue, we propose to discuss
a two-orbital tight-binding model with energy levels belonging to the 
$\Gamma_8$ quartet. The model reproduces the experimentally measured FSs well along the high-symmetry planes namely 
(100), (110) etc, which are part of the ellipsoid like three-dimensional FSs with
the squarish cross sections. With this realistic electronic structure, we examine the nature of of instability or 
fluctuations in the Hubbard-like model with standard onsite Coulomb interaction terms considered usually in a multiorbital system such as
iron-based superconductors. This is accomplished by studying behavior of the susceptibilities corresponding to the various 
multipolar moments.
\section{Model Hamiltonian}
Single particle state in the presence of strong spin-orbit coupling 
is defined by using total angular momentum ${{\bf j}} = \textit{{\bf l}} + \textit{{\bf s}}$, which yields 
low-lying sextet and high-lying octet 
for $j = 5/2$ and $7/2$, respectively in the case of $f$-electron with ${l} = 3$. Therefore, with the 
number of electrons $n$ being 1, it is the low lying 
sextet, which is relevant in the case of Ce$^{3+}$ ions. These ions are in the octahedral environment with 
corners being occupied by the six B ions. Therefore, the 
sextet is further split into $\Gamma_8$ quartet which forms the ground state of CeB$_6$ and a high lying
$\Gamma_7$ doublet separated by $\sim 500K$. $\Gamma_8$ quartet 
involves two Kramers doublet and each doublet can be treated 
as spin-$\frac{1}{2}$ system.\cite{hotta}

Using $\Gamma_8$ quartet, kinetic part of our starting Hamiltonian is 
\begin{equation}
 \mathcal{H}_0 = \sum_{{\bf i},{\bf j}}\sum_{\mu,\nu}\sum_{\sigma, {\sigma}^{\prime}} t_{{\bf i};{\bf j}}^{\mu 
 \sigma;\nu {\sigma}^{\prime}} 
(f_{{\bf i}\mu\sigma}^\dagger f_{{\bf j}\nu{\sigma}^{\prime}} + \text{H.c.}),
\end{equation} 
where $t_{{\bf i};{\bf j}}^{\mu 
 \sigma;\nu {\sigma}^{\prime}}$ are the hopping elements from orbital $\mu$ with psuedospin $\sigma$ at site 
 ${\bf i}$ to orbital $\nu$ with psuedospin $\sigma^{\prime}$ at site ${\bf j}$. The operator 
 $f_{{\bf i} \mu \sigma}^\dagger$ ($f_{{\bf i} \mu \sigma}$) creates (destroys) 
a $f$ electron in the $\mu$ orbital of site ${\bf i}$ with psuedo spin $\sigma$. These are given 
explicitly in terms of the $z$-components of the total
angular momentum $j = 5/2$ as follows,
\begin{eqnarray}
 f_{i1 \uparrow} &=& \sqrt{\frac{5}{6}}c_{i-\frac{5}{2}}+\sqrt{\frac{1}{6}}c_{i\frac{3}{2}} \nonumber\\
 f_{i1 \downarrow} &=& \sqrt{\frac{5}{6}}c_{i\frac{5}{2}}+\sqrt{\frac{1}{6}}c_{i-\frac{3}{2}} \nonumber\\
 f_{i2 \uparrow} &=& c_{i-\frac{1}{2}},\,\,\,f_{i2 \downarrow} = c_{i\frac{1}{2}}.
\end{eqnarray} 
As can be seen in Fig. \ref{orb}, $\Gamma_{8}$ orbitals are similar in structure to the $d$-orbitals $d_{x^2-y^2}$ 
and $d_{3z^2-r^2}$.

\begin{figure}
\begin{center}
\vspace*{-4mm}
\hspace*{-2mm}
\psfig{figure=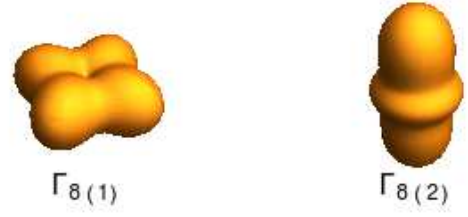,width=85mm,angle=0}
\vspace*{-2mm}
\end{center}
\caption{$\Gamma_{8(1)}$ and $\Gamma_{8(2)}$ orbitals with a similar structure as $d$-orbitals $d_{x^2-y^2}$ 
and $d_{3z^2-r^2}$, respectively.}
\label{orb}
\end{figure}  
\begin{figure}
\begin{center}
\vspace*{-8mm}
\hspace*{-8mm}
\psfig{figure=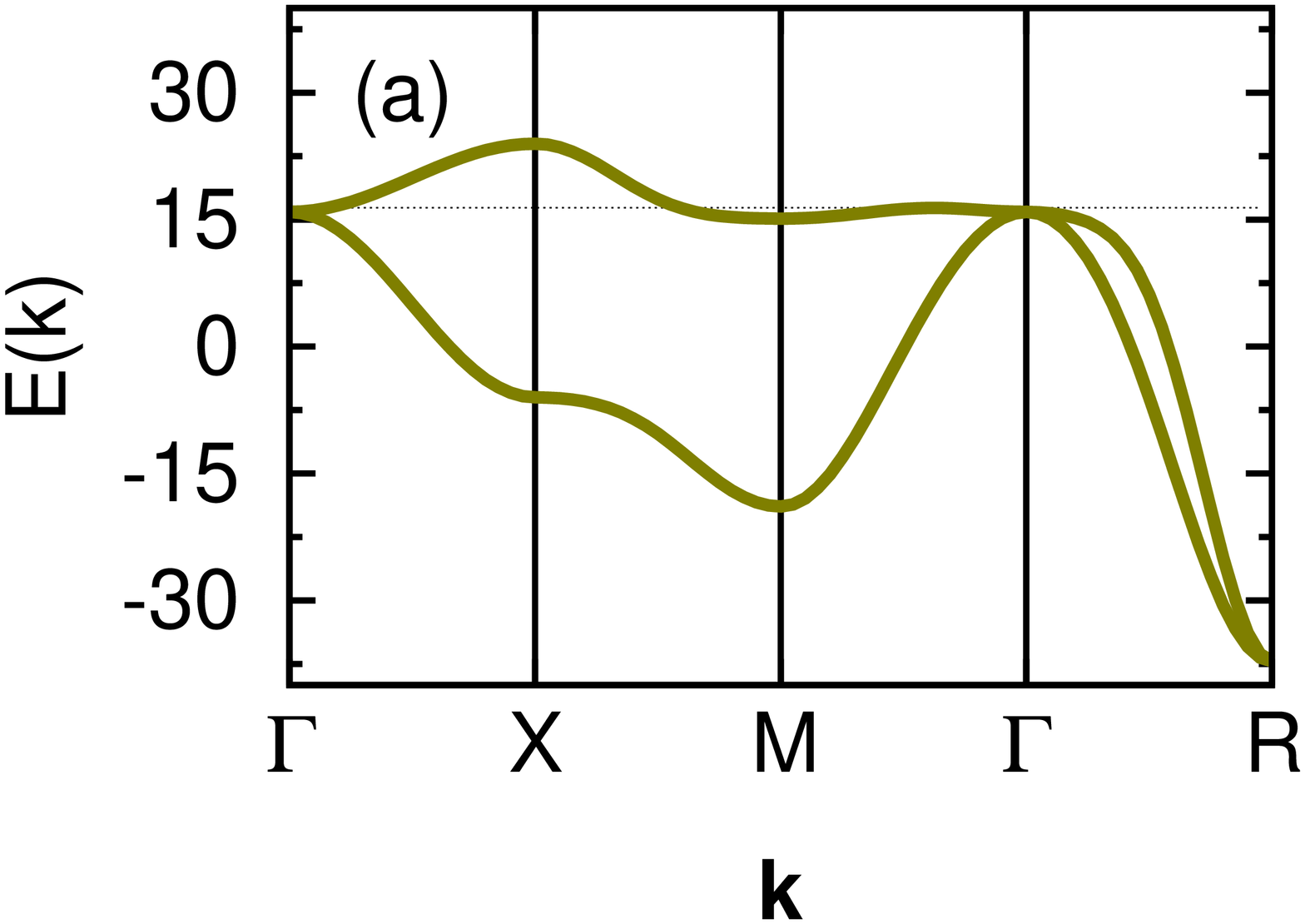,width=32.5mm,angle=-90}
\hspace*{-9mm}
\psfig{figure=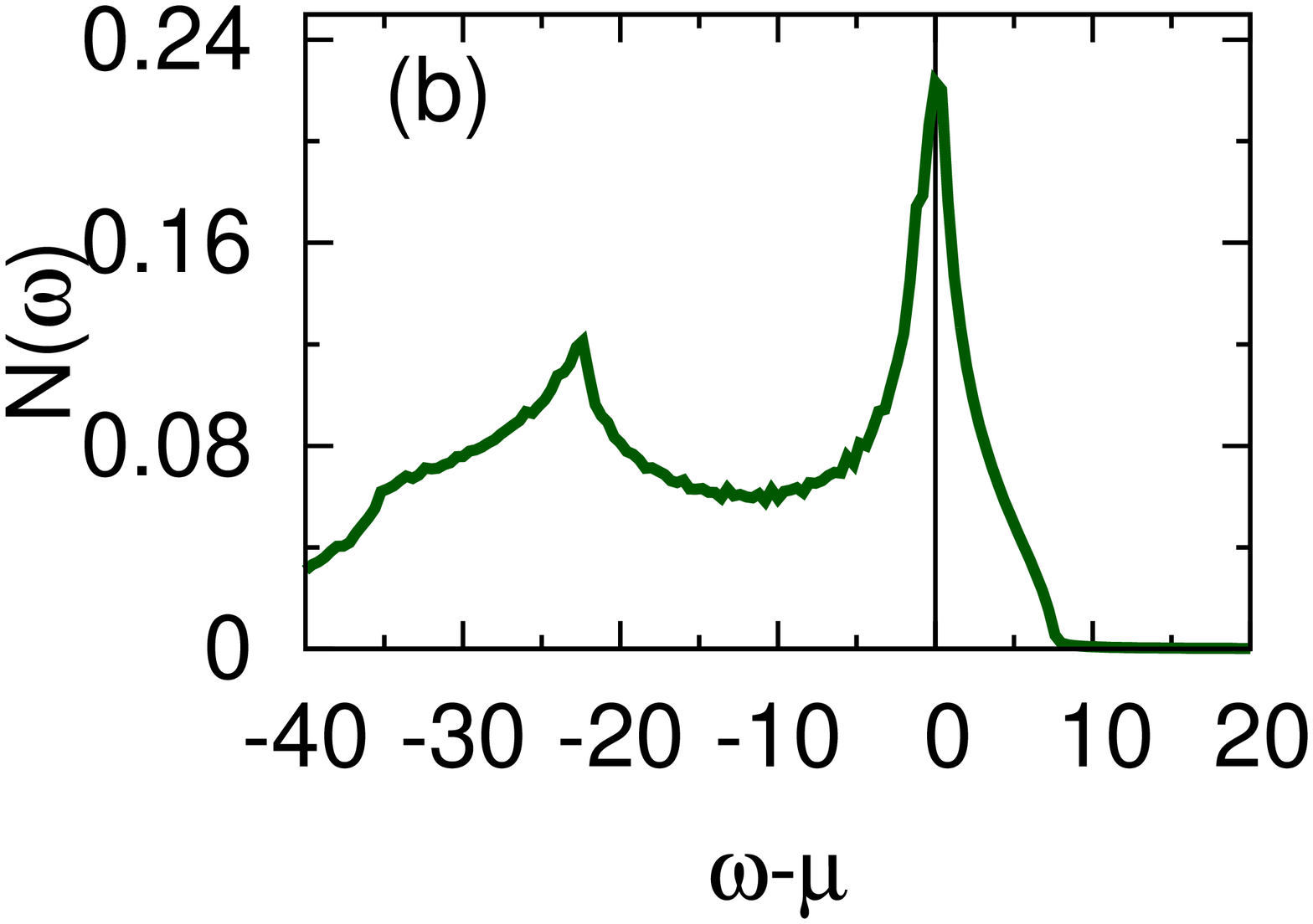,width=33.5mm,angle=-90}
\vspace*{-6mm}
\end{center}
\caption{(a) Electron dispersions along high-symmetry direction $\Gamma$-X-M-$\Gamma$-R and (b) DOS, which is peaked near 
the Fermi level.}
\label{qpi1}
\end{figure}  
\begin{figure}
\begin{center}
\vspace*{-10mm}
\hspace*{-2mm}
\psfig{figure=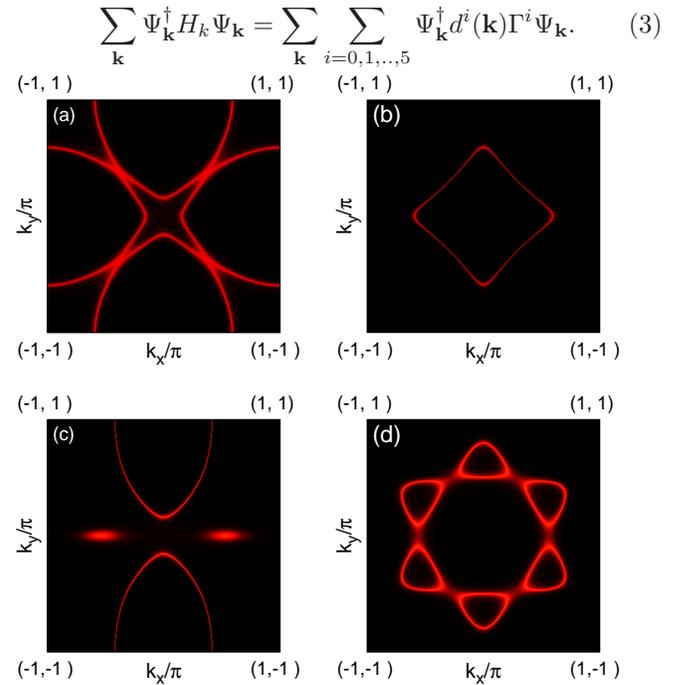,width=85mm,angle=0}
\vspace*{-4mm}
\end{center}
\caption{Electron dispersions along high-symmetry plane (a) (100), (b) at $\pi$ along 
(100), (c) (110) and (d) (111) obtained for the chemical potential $\mu$ = 16.4.}
\label{disp}
\end{figure}  
Kinetic energy after the Fourier transform can be expressed in terms of $\Gamma$ matrices defined as 
$\hat{\Gamma}^{0,1,2,3,4,5} = (\hat{\tau}_0 \hat{\sigma}_0,\hat{\tau}_z \hat{\sigma}_0,\hat{\tau}_x 
\hat{\sigma}_0,\hat{\tau}_y \hat{\sigma}_x,\hat{\tau}_y \hat{\sigma}_y,\hat{\tau}_y \hat{\sigma}_z)$,
where ${\sigma}_i$s and ${\tau}_i$s are Pauli's matrices corresponding to the spin and orbital degrees of freedom, respectively. So that
\begin{equation}
 \sum_{\k} \Psi^{\dagger}_{\k} H_k \Psi_{\k} = \sum_{\k} \sum_{i = 0,1,..,5} \Psi^{\dagger}_{\k} d^i (\k) \Gamma^{i}\Psi_{\k}.
\end{equation}
Here, $\Psi^{\dagger}_{\k} = (f^{\dagger}_{\k 1\uparrow},f^{\dagger}_
{\k 2\uparrow},f^{\dagger}_{\k 1\downarrow},f^{\dagger}_{\k 2\downarrow})$ is the electron field with $d^i (\k)$s  
\begin{eqnarray}
 d^0(\k) &=& -\mu+8t\phi_{0}(\k)+\frac{28}{3}t^{\prime}\phi^{\prime}_{0}(\k)+\frac{128}{9}t^{\prime \prime}
 \phi^{\prime \prime}_0(\k) \nonumber\\
 d^1(\k) &=& 4t\phi_1(\k)-\frac{2}{3}t^{\prime}\phi^{\prime}_1(\k) \nonumber\\
 d^2(\k) &=& -4\sqrt{3}t\phi_2(\k)+\frac{2}{\sqrt{3}}t^{\prime}\phi^{\prime}_2(\k) \nonumber\\
 d^3(\k) &=& \frac{16}{\sqrt{3}}t^{\prime}\phi^{\prime}_3(\k)+\frac{128}{9\sqrt{3}}t^{\prime \prime}
 \phi^{\prime \prime}_3(\k) \nonumber\\
 d^4(\k) &=& \frac{16}{\sqrt{3}}t^{\prime}\phi^{\prime}_4(\k)+\frac{128}{9\sqrt{3}}
 t^{\prime \prime}\phi^{\prime \prime}_4(\k) \nonumber\\
 d^5(\k) &=& \frac{16}{\sqrt{3}}t^{\prime}\phi^{\prime}_5(\k)+\frac{128}{9\sqrt{3}}
 t^{\prime \prime}\phi^{\prime \prime}_5(\k).
\end{eqnarray}
$t^{\prime}$ and $t^{\prime \prime}$ are the second and third next-nearest
neighbor hoping parameters. Various $\phi(\k)$s are expressed in terms of cosines and sines of the 
components of momentum in the Brillouin zone as
\begin{eqnarray}
      \phi_0&=&\cos k_x+\cos k_y+\cos k_z  \nonumber\\
      \phi^{\prime}_0&=&\cos k_y \cos k_z+\cos k_z \cos k_x+\cos k_x \cos k_y \nonumber\\
      \phi_1&=&\cos k_x+\cos k_y-2 \cos k_z \nonumber\\
      \phi^{\prime}_1&=&\cos k_y \cos k_z+\cos k_z \cos k_x-2 \cos k_x \cos k_y \nonumber\\
      \phi_2&=&\cos k_x-\cos k_y \nonumber\\
      \phi^{\prime}_2&=&\cos k_y \cos k_z-\cos k_z \cos k_x \nonumber\\
      \phi^{\prime}_3&=&\sin k_y \sin k_z \nonumber\\
      \phi^{\prime}_4&=&\sin k_z \sin k_x \nonumber\\
      \phi^{\prime}_5&=&\sin k_x \sin k_y \nonumber\\
      \phi^{\prime \prime}_0 &=& \cos k_x \cos k_y \cos k_z \nonumber\\
      \phi^{\prime \prime}_3 &=& \cos k_x \sin k_y \sin k_z \nonumber\\
      \phi^{\prime \prime}_4 &=& \sin k_x \cos k_y \sin k_z \nonumber\\
      \phi^{\prime \prime}_5 &=& \sin k_x \sin k_y \cos k_z.
\end{eqnarray}

In the absence of the second and third nearest-neighbor hopping, the kinetic part of the 
Hamiltonian reduces to that of manganites\cite{dagotto} with the only difference of a constant multiplication factor.
In the following, the unit of energy is set to be $t$. Calculated electron 
dispersions for $t^{\prime}  =-0.38$ and $t^{{\prime}{\prime}} = 0.18$, which consists
of doubly degenerate eigenvalues, are shown in Fig. \ref{disp}(a) along the high symmetry directions. 
 A large hole pocket near X and the extrema exhibited
by two bands near $\Gamma$ just below the Fermi level are broadly in agreement with 4$f$ 
dominated part in the band-structure calculations. The density of states (DOS) show 
  two peaks with larger one being in the vicinity of the Fermi level (Fig. \ref{disp}(b)). It is not unexpected particularly because
  of the flatness of the 
  two bands near $\Gamma$ contributing mostly to the DOS at the Fermi level. Interestingly, a
  hot spot near $\Gamma$ has been observed also in the ARPES measurements, which points towards the possibility of 
 strong ferromagnetic fluctuations.\cite{neupane} Here, the chemical potential is chosen to be 16.4 to 
obtain a better agreement with the ARPES FSs. 
 \begin{figure}
\begin{center}
\vspace*{-4mm}
\hspace*{-2mm}
\psfig{figure=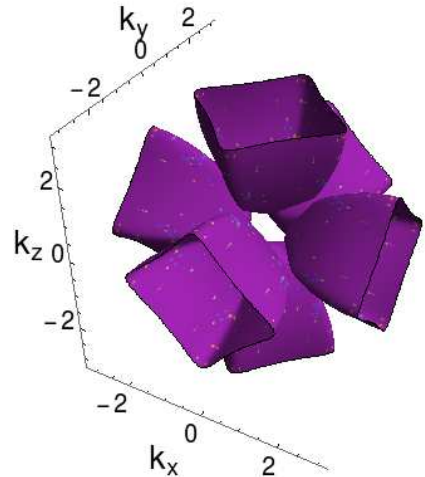,width=60mm,angle=0}
\vspace*{-4mm}
\end{center}
\caption{Fermi surfaces in the Brillouin zone for the chemical potential $\mu$ = 16.4.}
\label{qpi1}
\end{figure}  

Fig. \ref{disp} shows FSs cut along different high-symmetry planes. It has an ellipse-like structure with 
major axis aligned along $\Gamma$-X for the (100) plane while touching each other along $\Gamma$-M direction.
On the other hand, the parallel plane at (0, 0, $\pi$) consists of a single squarish pocket around that point. In the absence of four-fold rotation symmetry for the (110) plane, two large ellipse-like FSs 
surfaces are present with the major axes along $\Gamma$-Y direction while small pockets exist along $\Gamma$-X direction. The six-fold rotation symmetry is reflected by the six pockets 
along (111) plane. All of them are obtained from the FSs shown in the whole Brillouin zone as in the Fig. 3. 
It consists of an ellipsoid-like FSs with largest semi-principal axes coinciding with $\Gamma$-X,
however, with a squarish cross section. An overall good 
agreement exists with the several recent ARPES
measurements.\cite{koitzsch,neupane} ARPES estimates are believed to more reliable when compared with the earlier
estimates from dHvA experiments carried out in the presence of magnetic field as 
the latter has the potential to affect the hot spots.
\section{Multipolar susceptibilities} 
\begin{figure*}
\begin{center}
\vspace*{-4mm}
\hspace*{-8mm}
\psfig{figure=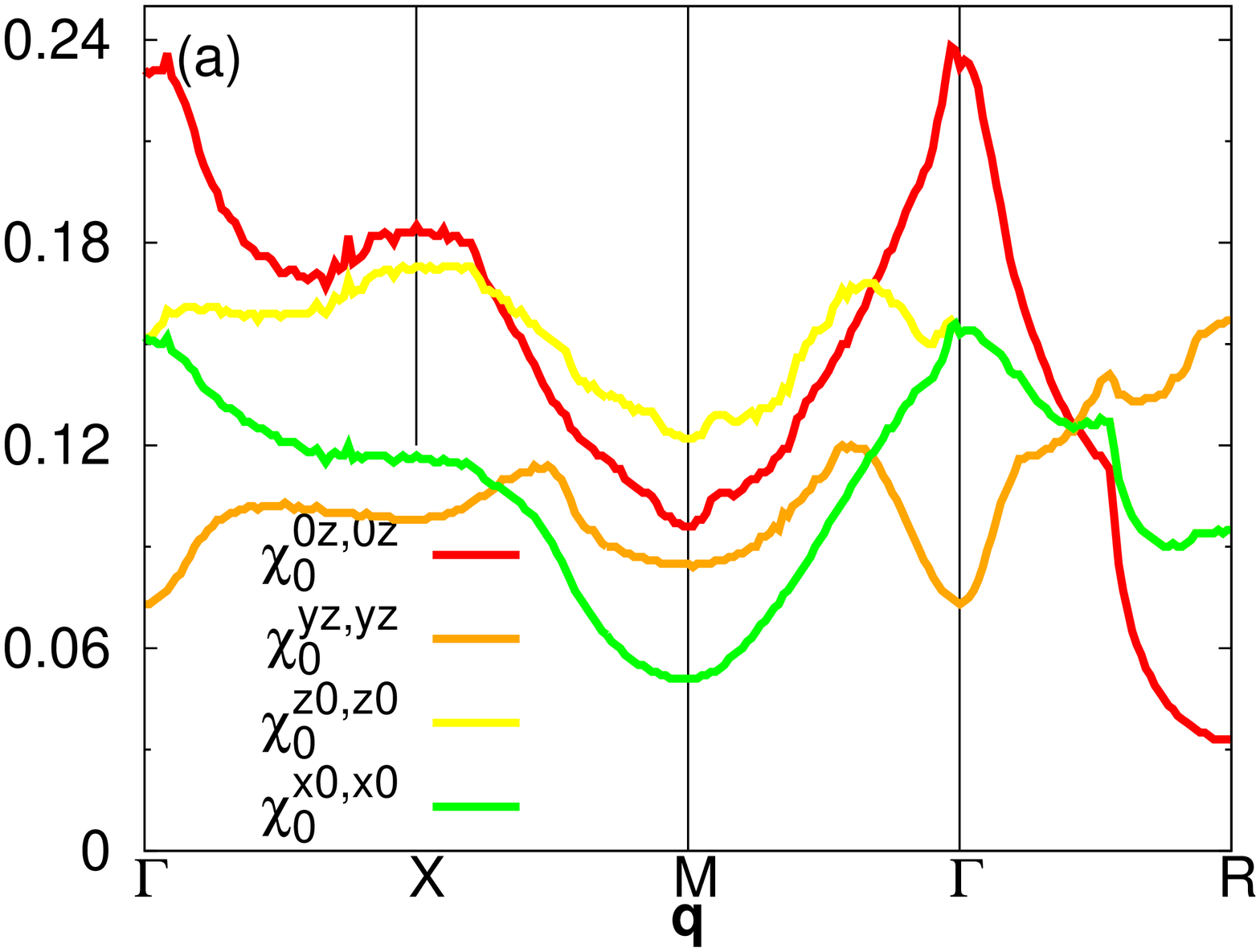,width=48mm,angle=-90}
\hspace*{0mm}
\psfig{figure=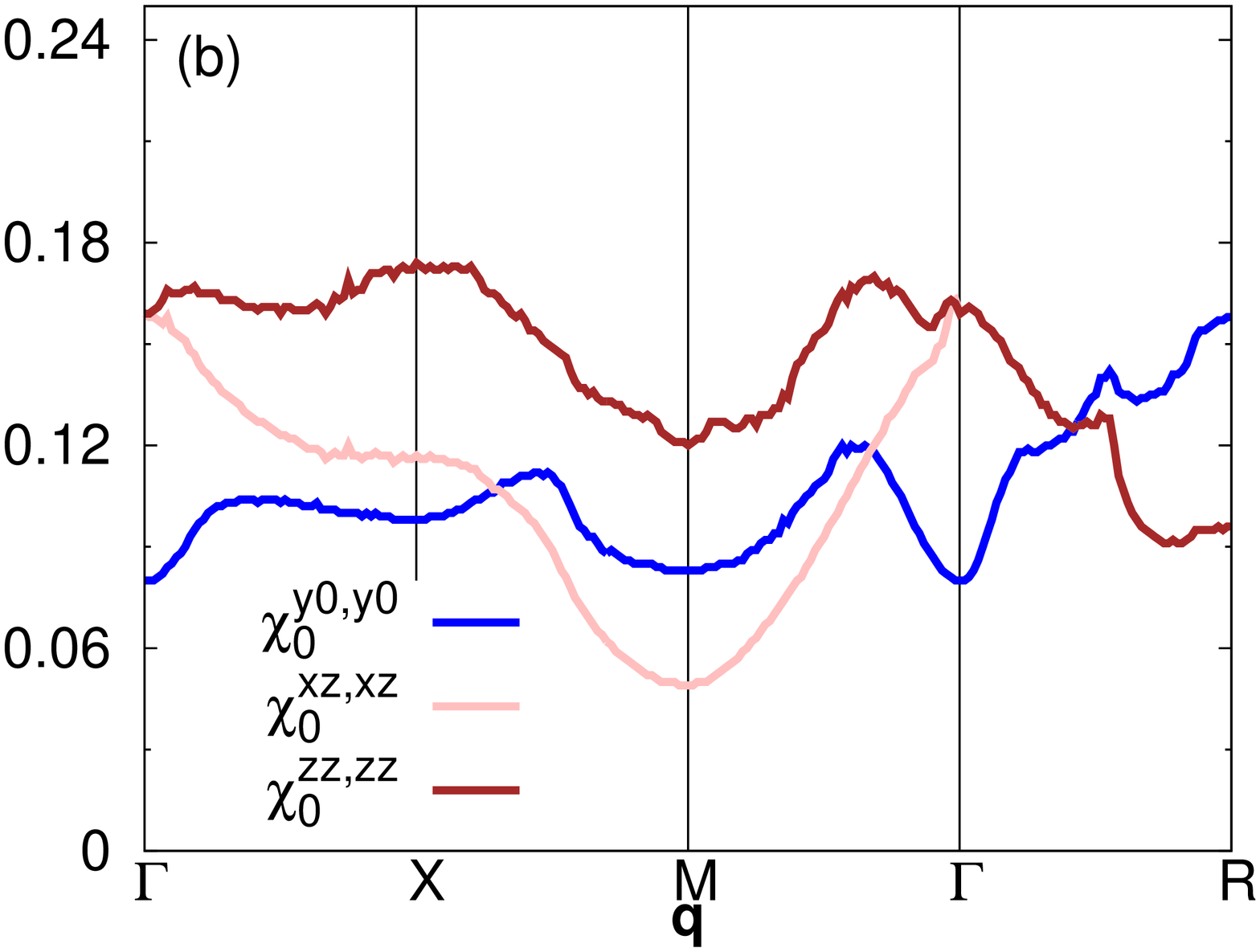,width=48mm,angle=-90}
\vspace*{-4mm}
\end{center}
\caption{(a) Magnetic and quadrupolar static susceptibilities 
along high-symmetry direction $\Gamma$-X-M-$\Gamma$-R. (b) Octapolar static susceptibilities.}
\label{mpi1}
\end{figure*}  
\begin{figure*}
\begin{center}
\vspace*{-4mm}
\hspace*{-8mm}
\psfig{figure=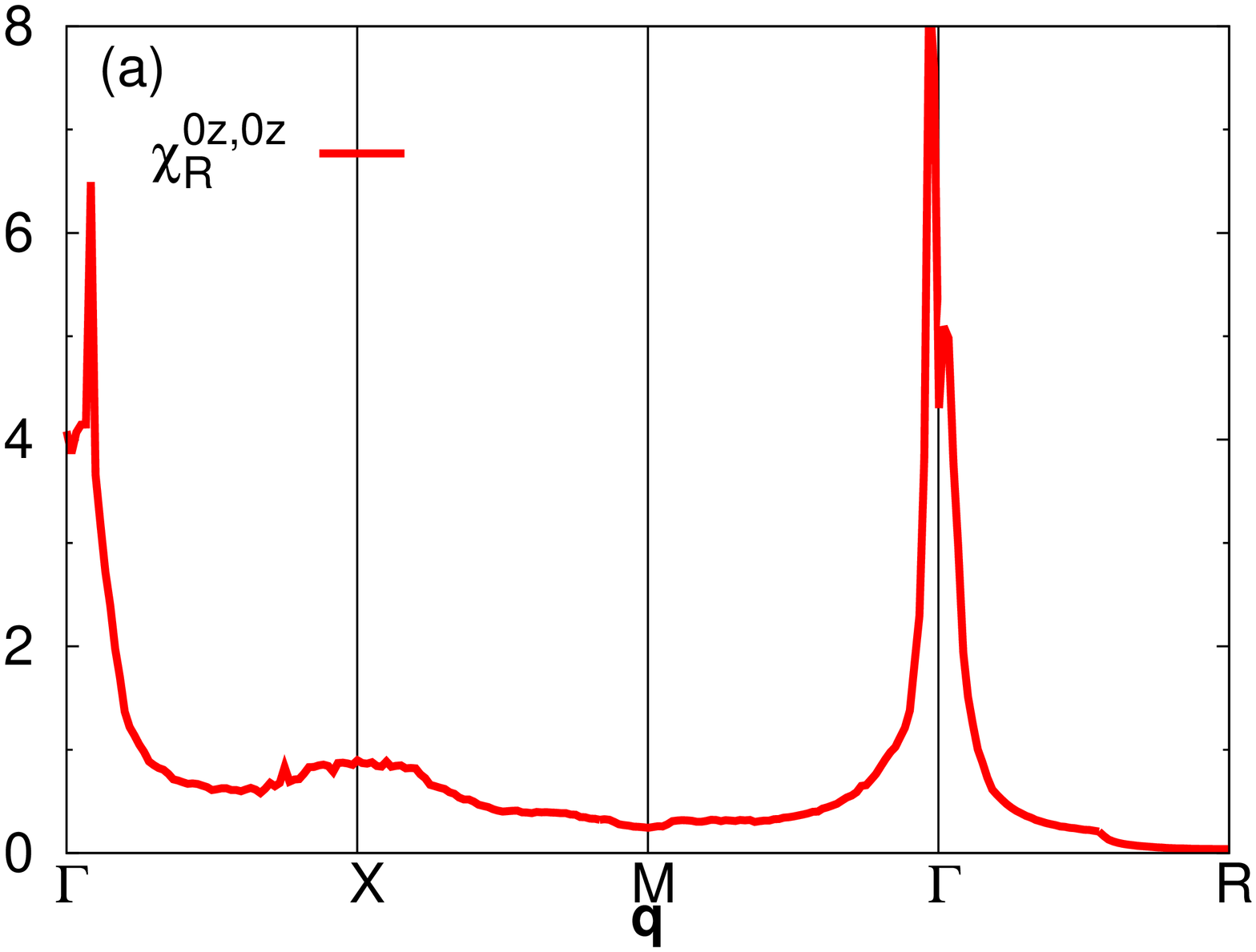,width=48mm,angle=-90}
\hspace*{0mm}
\psfig{figure=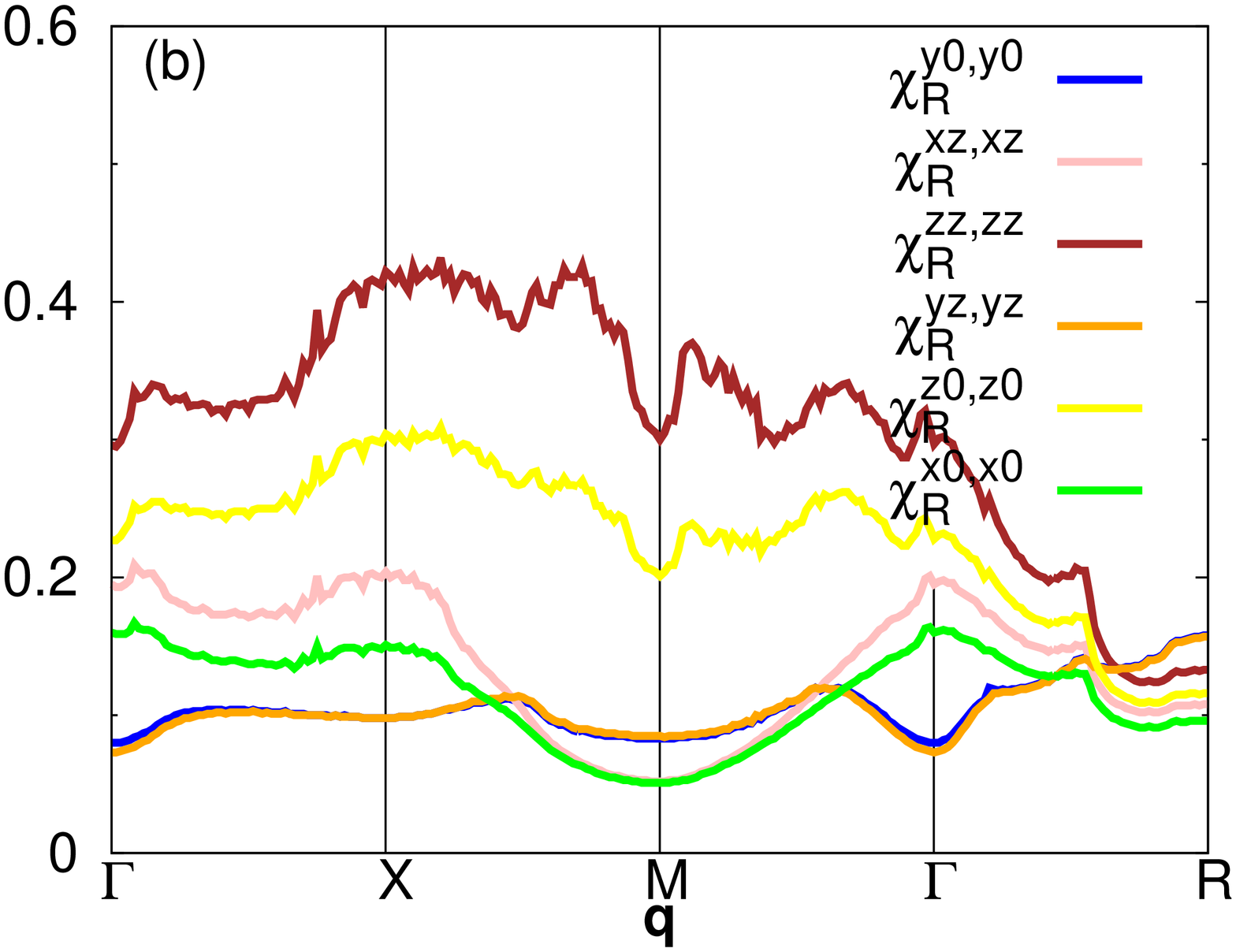,width=48mm,angle=-90}
\vspace*{-4mm}
\end{center}
\caption{Multipolar static susceptibilities calculated at the RPA-level for $U = 15$ and $J = 0.16U$. (a) Spin susceptibility 
diverges near (0, 0, 0). (b) For the same set of interaction parameters quadrupolar and octapolar susceptibilities are divergenceless.}
\label{mpi2}
\end{figure*}  
Sixteen multipolar moments can be defined for
the $\Gamma_8$ state including one charge, three dipole,
five quadrupole and seven octapole, which are rank-0,
rank-1, rank-2 and rank-3 tensors, respectively. The dipole belongs to the $\Gamma^{-}_4$ irreducible representation, where 
$-$ sign denote the breaking of time reversal symmetry. It's components are given by the outer product of Pauli's
matrices $\hat{\tau_0}\hat{\sigma_i}$s. The 
quadrupole moments belonging to $\Gamma^{+}_3$ are $\hat{\tau_x}\hat{\sigma_0}$ and $\hat{\tau_z}\hat{\sigma_0}$ while
those belonging to $\Gamma^{+}_5$ irreducible representations are expressed as $\hat{\tau_i}\hat{\sigma_y}$s. The 
octapular moments with $\Gamma^{-}_2$ representation is $\hat{\tau_y}\hat{\sigma_0}$, whereas the $z$-component of 
those belonging to $\Gamma^-_4$ and $\Gamma^-_5$ are $2\hat{\tau_z}\hat{\sigma_z}$ and $2\hat{\tau_x}\hat{\sigma_z}$, 
respectively. 

In order to examine the multipolar ordering instabilities, 
we calculate susceptibilities while considering only the $z$-component whenever component along
three coordinate axes are present as that will be sufficient because of the cubic symmetry. Multipolar 
susceptibilities are defined as\cite{takimoto1}
\be
{\chi}^{pq,rs}(\q,i\omega_n)= \frac{1}{\beta} \int^{\beta}_0{d\tau e^{i \omega_{n}\tau}\langle T_\tau
[{\cal O}^{pq}_{\q}(\tau) {\cal O}^{rs}_{ -\q}(0)]\rangle},
\ee
where 
\be
{\cal O}^{pq}_{\q}= \sum_{\bf k} \sum_{\sigma \sigma^{\prime}} \sum_{\mu \mu^{\prime}} f^{\dagger}_{\mu \sigma}(\k+\q)
\tau^{p}_{\mu \mu^{\prime}} \sigma^{q}_{\sigma \sigma^{\prime}} f_{\mu^{\prime} \sigma^{\prime}}(\k).
\ee
They can be expressed in terms of 
\begin{eqnarray}
\chi^{\sigma_1 \sigma_2;\sigma_4 \sigma_3}_{\mu_1 \mu_2;\mu_4 \mu_3}(\q,i\omega_n)
&=& \frac{1}{\beta} \int^{\beta}_0 d\tau \sum_{\k,\k^{\prime}}\langle T_{\tau} f^{\dagger}_{\k+\q \mu_1 \sigma_1}(\tau)
f_{\k \mu_2 \sigma_2}(\tau)\nonumber\\
&\times&f^{\dagger}_{\k^{\prime}-\q^{\prime} \mu_3 \sigma_3}(0)f_{\k^{\prime} \mu_4 \sigma_4}(0)\rangle
\end{eqnarray}
which form a 16$\times$16 matrix. Thus, the dipole or spin susceptibility is given by
\be
{\chi}^{0z,0z}(\q,i\omega_n)= \sum_{\sigma \sigma^{\prime}} \sum_{\mu \mu^{\prime}} \sigma \sigma^{\prime}\chi^{\sigma \sigma;\sigma^{\prime}\sigma^{\prime}}
_{\mu \mu;\mu^{\prime} \mu^{\prime}}(\q,i\omega_n),
\ee
where $\sigma$ and $\mu$ in front of $\chi$ takes +1 or -1 corresponding to the two spin or orbital degrees of freedom.
Various quadrupolar and octapolar susceptibilities are given as
\begin{eqnarray}
{\chi}^{x0,x0}(\q,i\omega_n) &=& \sum_{\sigma } \sum_{\mu \mu^{\prime}}
\chi^{\sigma \sigma;\sigma \sigma}_{\mu \bar{\mu};\mu^{\prime} \bar{\mu}^{\prime}}(\q,i\omega_n) \nonumber\\
{\chi}^{z0,z0}(\q,i\omega_n) &=& \sum_{\sigma } 
\sum_{\mu \mu^{\prime}} \mu \mu^{\prime}\chi^{\sigma \sigma;\sigma\sigma}
_{\mu \mu;\mu^{\prime} \mu^{\prime}}(\q,i\omega_n) \nonumber\\
{\chi}^{yz,yz}(\q,i\omega_n) &=& -i^2\sum_{\sigma \sigma^{\prime}} \sum_{\mu \mu^{\prime}}
\sigma \sigma^{\prime}\mu \mu^{\prime}\chi^{\sigma \sigma;\sigma^{\prime}\sigma^{\prime}}
_{\mu \bar{\mu};\mu^{\prime} \bar{\mu}^{\prime}}(\q,i\omega_n) \nonumber\\
\end{eqnarray} 
and 
\begin{eqnarray}
 {\chi}^{y0,y0}(\q,i\omega_n) &=& -i^2\sum_{\sigma } \sum_{\mu \mu^{\prime}} \mu \mu^{\prime}
\chi^{\sigma \sigma;\sigma \sigma}_{\mu \bar{\mu};\mu^{\prime} \bar{\mu}^{\prime}}(\q,i\omega_n) \nonumber\\
{\chi}^{xz,xz}(\q,i\omega_n) &=& \sum_{\sigma \sigma^{\prime}} 
\sum_{\mu \mu^{\prime}} \sigma \sigma^{\prime} \chi^{\sigma \sigma;\sigma^{\prime}\sigma^{\prime}}
_{\mu \bar{\mu};\mu^{\prime} \bar{\mu}^{\prime}}(\q,i\omega_n) \nonumber\\
{\chi}^{zz,zz}(\q,i\omega_n) &=& \sum_{\sigma \sigma^{\prime}} \sum_{\mu \mu^{\prime}}
\sigma \sigma^{\prime}\mu \mu^{\prime}\chi^{\sigma \sigma;\sigma^{\prime}\sigma^{\prime}}
_{\mu \mu;\mu^{\prime} {\mu}^{\prime}}(\q,i\omega_n),
\end{eqnarray} 
respectively.

Fig. 5 shows different static multipolar susceptibilities with well-defined peaks for some while broad hump like structure for the 
other. Particularly, the spin susceptibility $\bar{\chi}^{0z,0z}$ is, among all, sharply peaked, however, 
at $\approx$ $\Q_3$ = (0, 0, 0). Quadrupolar susceptibility $\bar{\chi}^{yz,yz}$ corresponding to the AFQ order observed 
 in experiments, on the other hand, does shows a peak near $\Q_1$. Other quadrupolar susceptibility $\bar{\chi}^{x0,x0}$
 is peaked near $\approx$ $\Q_3$ while $\bar{\chi}^{z0,z0}$ has 
a broad hump like structure near ($\pi, 0, 0$) and a peak slightly away from ($\pi/2, \pi/2, \pi/2$). 
We further note that $\bar{\chi}^{x0,x0} = \bar{\chi}^{xz,xz}$, 
$\bar{\chi}^{y0,y0} = \bar{\chi}^{yz,yz}$ and $\bar{\chi}^{z0,z0} = \bar{\chi}^{zz,zz}$ as shown in Fig. 5(b)
\section{Multipolar susceptibilities in the presence of interaction}
In order to investigate the role of electron-electron correlation, we consider the standard onsite 
Coulomb interaction terms given as
\begin{eqnarray}
\mathcal{H}_{int} &=& U \sum_{{\bf i},\mu} n_{{\bf i}\mu \uparrow} n_{{\bf i}\mu \downarrow} + (U' -
\frac{J}{2}) \sum_{{\bf i}, \mu<\nu} n_{{\bf i} \mu} n_{{\bf i} \nu} \nonumber \\ 
&-& 2 J \sum_{{\bf i}, \mu<\nu} {\bf{S_{{\bf i} \mu}}} \cdot {\bf{S_{{\bf i} \nu}}} + J \sum_{{\bf i}, \mu<\nu, \sigma} 
f_{{\bf i} \mu \sigma}^{\dagger}f_{{\bf i} \mu \bar{\sigma}}^{\dagger}f_{{\bf i} \nu \bar{\sigma}}
f_{{\bf i} \nu \sigma}, \nonumber\\
\label{int}
\end{eqnarray}
in a manner similar to the various correlated multiorbital systems. First term represents the
intraorbital Coulomb interaction for each orbital. Second and third term 
represent the density-density interaction and Hund's coupling between the two orbitals. Fourth term represents the 
pair-hopping energy whereas the 
condition $U^{\prime}$ = $U$ - $2J$ is essential for the rotational invariance.

Multipolar susceptibilities in the presence of interaction can be obtained from Dyson's equation yielding 
\be
\hat{\chi}_R(\q, i\omega) = (\hat{{\bf 1}}-\hat{U}\hat{{\chi}}(\q, i\omega))^{-1}\hat{{\chi}}(\q, i\omega).
\ee
Here, $\hat{{\bf 1}}$ is a $16\times16$ identity matrix, whereas the interaction matrix is given by\cite{scherer}
\begin{eqnarray}
&&{U}^{\sigma_1 \sigma_2;\sigma_3 \sigma_4}_{\mu_1 \mu_2;\mu_3 \mu_4}  \nonumber\\
&&= \left\{
\begin{array}{@{\,} l @{\,} c}
-U & (\mu_1=\mu_2=\mu_3=\mu_4, \sigma_1=\sigma_2\ne\sigma_3=\sigma_4)\\
-U^{\prime} & (\mu_1=\mu_2\ne\mu_3=\mu_4, \sigma_1=\sigma_2\ne\sigma_3=\sigma_4)\\
-J & (\mu_1=\mu_4\ne \mu_2=\mu_3,\sigma_1=\sigma_2\ne\sigma_3=\sigma_4)\\
-J^{\prime} & (\mu_1=\mu_3\ne \mu_2=\mu_4,\sigma_1=\sigma_2\ne\sigma_3=\sigma_4)\\
-(U-J^{\prime})& (\mu_1=\mu_2\ne\mu_3=\mu_4, \sigma_1=\sigma_2=\sigma_3=\sigma_4)\\
(U-J^{\prime}) & (\mu_1=\mu_4\ne\mu_2=\mu_3, \sigma_1=\sigma_2=\sigma_3=\sigma_4)\\
U & (\mu_1=\mu_2 = \mu_3=\mu_4,\sigma_1=\sigma_4\ne\sigma_2=\sigma_3)\\
U^{\prime} & (\mu_1=\mu_4 \ne \mu_2=\mu_3,\sigma_1=\sigma_4\ne\sigma_2=\sigma_3)\\
J & (\mu_1=\mu_2\ne \mu_3=\mu_4,\sigma_1=\sigma_4\ne\sigma_2=\sigma_3)\\
J^{\prime} & (\mu_1=\mu_3\ne \mu_2=\mu_4,\sigma_1=\sigma_4\ne\sigma_2=\sigma_3)\\
0 & (\mathrm{otherwise})
\end{array}\right. . \nonumber\\
\end{eqnarray}

Fig. \ref{mpi2} show the multipolar static susceptibilities at the RPA-level. As expected, the RPA
spin susceptibility requires the smallest critical interaction strength $U = 15$ ($U$/$W$ $< 1/3$) with $J = 0.16U$
to show the divergence. Interestingly, it diverges near $\Q_3$ instead of at the
AFM ordering wave vector $\Q_2$, which is not surprising
 because there exists a large DOS near $\Gamma$ that leads also to the peak near $\Q_3$ in bare spin suspceptibility. 
 Thus, AFQ instability corresponding to the $\Gamma^+_5$ representation is absent in the model despite the bare quadurpole susceptibility being peaked near 
 $\Q_1$. However, we believe that the strong low-energy ferromagnetic fluctuations in the paramagnetic phase may have 
 important implications for the persistent ferromagnetic correlations in various ordered phases as observed by various experiments.\cite{demishev2,jang}
  \section{Conclusions and discussions}
 In conclusions, we have described a tight-binding model with the bases as $\Gamma_8$,
 which captures the salient features of 
 the Fermi surfaces along the high-symmetry planes as observed in the ARPES measurements. A large density of state is
 obtained near the Fermi level due to the flatness of the bands close to $\Gamma$, which bears a remarkable similarity to the 
 hot-spot observed in another ARPES experiments. Multipolar susceptibilities calculated with the 
 standard onsite Coulomb interactions as in other multiorbital systems show that it is the spin
 susceptibility that exhibits strongest diverging behavior. Moreover, it does so in the low-momentum region implying 
 an underlying ferromagnetic instability.
 
  It is clear that nature of the instability obtained with the 
  realistic electronic structure is different from the actual order in CeB$_6$. However, it is important to note that some of the recent experiments have provided 
  the evidence of strong ferromagnetic correlations in the ordered phases. For instance, there exists magnetic 
 spin resonance in the AFQ phase, which has been attributed to the FM correlations. Further, the most intense
 spin-wave excitation modes have been observed at zero-momentum instead of the AFM ordering wavevector by the
 INS measurements in the coexistence phase, which continues to be present even in the 
 AFQ phase. A similar INS measurement in the paramagnetic phase is highly desirable to probe the existence of 
 ferromagnetic correlations in the paramagnetic phase. So far only an indirect indication in the form of 
 hot-spot observed by ARPES near $\Gamma$ is available. In order 
  to understand above mentioned features, we believe that the 
  strong low-energy ferromagnetic fluctuations obtained within the two-orbital model with the 
  realistic electronic structure may be an important step. To explain AFQ and other multipole order, 
 it would perhaps be necessary to include the local-exchange terms involving AFQ and multipolar moments. Such a proposal should be the subject matter of 
 future investigation in order to describe various complex ordering phenomena as well as associated unusual features within 
 a single model.
 
We acknowledge the use of HPC clusters at HRI.

\end{document}